\documentstyle[preprint,prd,aps,epsf]{revtex}
\begin{document}
\draft
\preprint{July 24, 1998}
\title{Quantum energy flow, dissipation and decoherence 
in mesoscopic dielectric structures}
\author{M. P. Blencowe}
\address{The Blackett Laboratory, Imperial College, London SW7~2BZ}

\maketitle

\begin{abstract}

We first present a summary of recent results concerning the phononic energy 
transport
properties of mesoscopic, suspended dielectric wires. We then discuss some
related open problems concerning the fundamental lower limits on the vibrational
damping rates of submicron-sized cantilever structures and also the possibility
to create and detect quantum superpositions of spatially separated states for
such structures.    
\vskip .75cm  
\noindent Keywords: quantum wires, Landauer thermal conductance,  
mesoscopic mechanical dissipation, quantum decoherence
\vskip 2.5cm
\noindent Address for correspondence: M. P. Blencowe, The Blackett Lab., Imperial
College, London SW7 2BZ, England. Fax: +44-171-594-7604. E-mail:
m.blencowe@ic.ac.uk   
\end{abstract}
\vskip 2cm

Recent experiments by Michael Roukes and coworkers on heat transport in 
submicron-sized, suspended dielectric wires \cite{tighe} have opened up new 
avenues for investigation in the fields of mesoscopics and phonon physics. 
Just as for electrons in conventional quantum wells, wires etc.,  phonons 
confined to the suspended wires can exhibit manifestly nonclassical behaviour. 
For example, the low temperature thermal conductance is quantized
in {\it universal} units $\pi k^2_{B}T/6\hbar$, analogous to the 
well-known $2e^2/h$ electronic conductance quantum 
\cite{angelescu,rego,blencowe}. This result follows from the Landauer formula
for the thermal conductance:
\begin{equation}
\kappa=\frac{\pi k^2_{B} T}{6\hbar}\sum_{n, n'}
\int_{\frac{E_{n,0}}{k_{B}T}}^{\infty}
d\epsilon\ g(\epsilon) T^{21}_{n' n}(\epsilon k_{B} T),
\label{thermalconductance}
\end{equation} 
where $T$ is the average temperature of the two phonon reservoirs (with
the reservoir temperature difference supposed small), $T^{21}_{n' n}(E)$
is the probability for a phonon with energy $E$ in subband $n$ of lead 1 to 
be (elastically) transmitted into  subband $n'$ of lead 2, and $E_{n,0}$ 
is the zone-centre 
energy of subband $n$. The function $g(\epsilon)$ is defined as follows:   
\begin{equation}
g(\epsilon)=\frac{3\epsilon e^{\epsilon}}{\pi^2 (e^{\epsilon}-1)^2}
\label{gfuncdef}
\end{equation}
and satisfies $\int_0^{\infty}d\epsilon\ g(\epsilon) =1$. Thus, 
in the absence of scattering, i.e. $T^{21}_{n' n}=\delta_{n' n}$,
a given subband $n$ contributes to the 
reduced conductance $\kappa/T$ the {\it universal} quantum 
$\pi k^2_{B}/6\hbar
\approx 9.465\times 10^{-13}\ {\rm W K}^{-2}$
in the limit $E_{n, 0}/k_{B} T \rightarrow 0$.
In Fig.\ \ref{fig1}, we show the temperature dependence of the reduced thermal
conductance for perfect GaAs wires with uniform, rectangular cross sections of
various dimensions comparable to those used in the experiments of 
Ref.\ \cite{tighe}. The only 
wire characteristics which are needed in order to determine the 
conductance are the zone-center frequencies $\omega_{n,0}$. 
These can be calculated  using the elegant numerical method developed in 
Ref.\ \cite{nishiguchi}. Note that there are no steplike features, a 
consequence of the broadness of the Bose-Einstein distribution as compared
with the  zone-centre energy separations.
There is, however, a plateau for $T\rightarrow 0$ where only phonons in the 
lowest
subband with $E_{n,0}=0$ contribute.
The  plateau has the value
four in universal  quantum units, a consequence of there being four
basic mode types: dilatational, torsional and two types of flexural 
mode.

The devices of Roukes and coworkers are in principle capable of measuring
much more than just the thermal conductance. Because of the very small volume
and hence heat capacity of the integrated electron gas thermometers, 
adsorption and emission of {\it single} phonons can produce {\it measurable}
fluctuations in the gas temperature.
From the magnitude of a given temperature fluctuation,
the energy of the absorbed or emitted phonon is known and, thus, there
is the possibility for high resolution phonon spectroscopy, one of the 
``holy grails" of the phonon physics community. 
In particular, by measuring the
fluctuation statistics for the absorption of phonons in a given narrow
energy interval, information can be obtained concerning the energy dependence 
of the phonon transmission probability for the suspended wires 
and also of the electron-phonon interaction in the wires. 
An initial discussion can be found in Ref. \cite{blencowe}.

We finish this note with a brief mention of two open problems which are somewhat 
related to the above discussion. The first concerns the vibrational damping 
rates of submicron-sized, single crystal cantilevers. In such small structures,
volume defects are rare and preliminary experiments suggest that excitation
of surface defects provides the dominant damping mechanism 
\cite{harrington,yasumura}. However, the presence and nature of surface 
defects is strongly dependent on the cantilever fabrication methods, so that 
appropriate modifications in these methods will lead to reduced surface
defect densities. A natural  question then is the following: 
what is the lowest possible damping rate which can be achieved for a 
given cantilever vibrational mode? In a defect-free, single crystal dielectric 
cantilever, only two dissipation mechanisms remain: 1) the coupling
of the cantilever vibrational modes to the substrate modes at the
cantilever base, and 2) the coupling  via the anharmonic interaction
 of the vibrational modes
 to the background
 thermal
phonon distribution in the cantilever. 
These mechanisms are unavoidable and thus set
a fundamental lower limit on the damping rate.
Submicron-sized cantilevers can have fundamental
flexural frequencies ranging from KHz to GHz and it may be the case 
that the relative  strengths of the two mechanisms varies considerably over 
this range. Just as for the energy flow properties, it is expected that 
submicron-sized structures with dimensions comparable to the thermal phonon 
wavelength will exhibit qualitatively different fundamental damping behaviour 
from that of bulk mechanical oscillators. It would be interesting to determine
the relative strengths of the fundamental damping mechanisms. They are 
well-characterised at the microscopic level and so should be accessible to
theoretical analysis.

The second problem concerns the 
possibility of creating and detecting quantum  superpositions
of spatially separated states for submicron-sized vibrating cantilevers.
In the field of quantum optics, Haroche and coworkers have successfully
demonstrated the creation and detection of superpositions of quantum states 
involving radiation fields \cite{brune}. A recent theoretical 
investigation by the Quantum Optics Group at Imperial College \cite{bose}
has suggested that 
such field states can be used to drive a 
submicron-sized, moveable mirror into a quantum superposition of spatially 
separated states. Such a quantum superposition will typically decohere
extremely rapidly into a classical mixture state as a result of dissipation.             
However, their estimates suggest that the decoherence
rates of submicron-sized moveable mirrors fabricated using similar methods to 
those of Roukes' group will be sufficiently low so as to be able to 
observe the  quantum  superpositions. 
In the light of this possibility, a natural question for a condensed matter 
physicist concerns whether an {\it electron} current  could be
used to drive a submicron-sized cantilever into a nonclassical state. 
One possible device is shown in Fig.\ \ref{fig2}. A moveable
quantum dot is located between two quantum wire contacts. Nanometer-scale 
gaps separate the quantum dot from the wire contacts, so that an electron can 
only move from one contact to the other via the quantum dot by tunneling 
across the gaps. When an electron is in the quantum dot it can impart 
some of its momentum to the dot, causing it to move, 
whereas when the electron is in one of the contacts it has no influence on 
the dot motion. Thus, the quantum uncertainty in a tunneling electron's 
location will induce some uncertainty in the dot's motion. It would be 
interesting to determine possible signatures in the measured electron
current of this quantum uncertainty in the dot motion.

\acknowledgments

The author would like to thank the organizers of Phonons 98   
for providing the opportunity to present this work.
Funding by the EPSRC under Grant No. GR/K/55493 is also 
acknowledged.

\begin{figure}
\caption{Reduced thermal conductance in universal units 
versus temperature for perfect GaAs wires 
with uniform  rectangular cross sections $200\ {\rm nm}\times 400\ {\rm nm}$
(solid line), $200\ {\rm nm}\times 300\ {\rm nm}$ (dashed line)
and $200\ {\rm nm}\times 100\ {\rm nm}$ (dotted line).}
\label{fig1}
\end{figure}

\begin{figure}
\caption{Model single electron potential profile for a movable quantum dot located
between two quantum wire contacts.} 
\label{fig2}
\end{figure}

\setcounter{figure}{0}
\begin{center}
\begin{figure}
\vskip 7in
\epsfxsize=6in
\epsffile{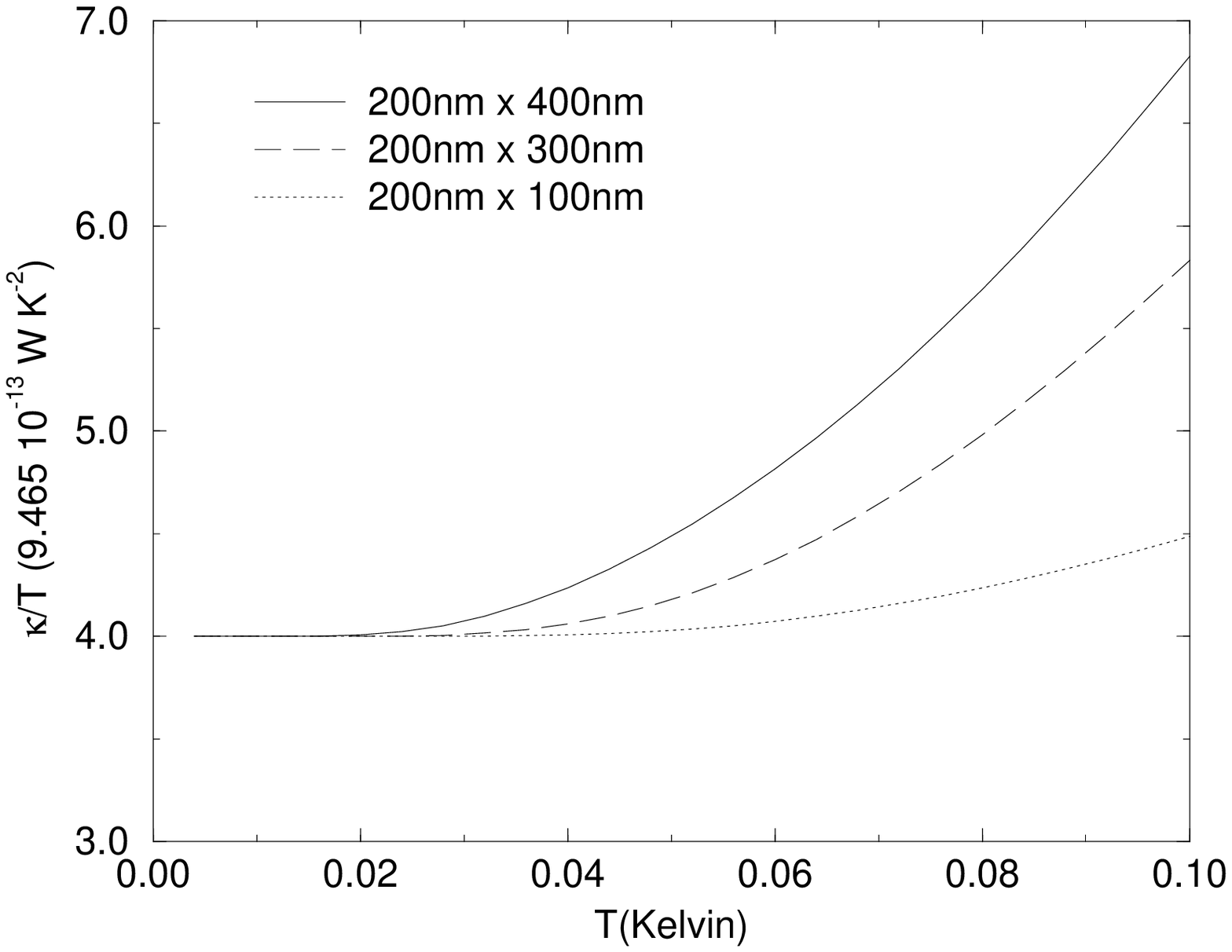}
\caption{}
\end{figure}
\end{center}

\begin{center}
\begin{figure}
\epsfxsize=6in
\epsffile{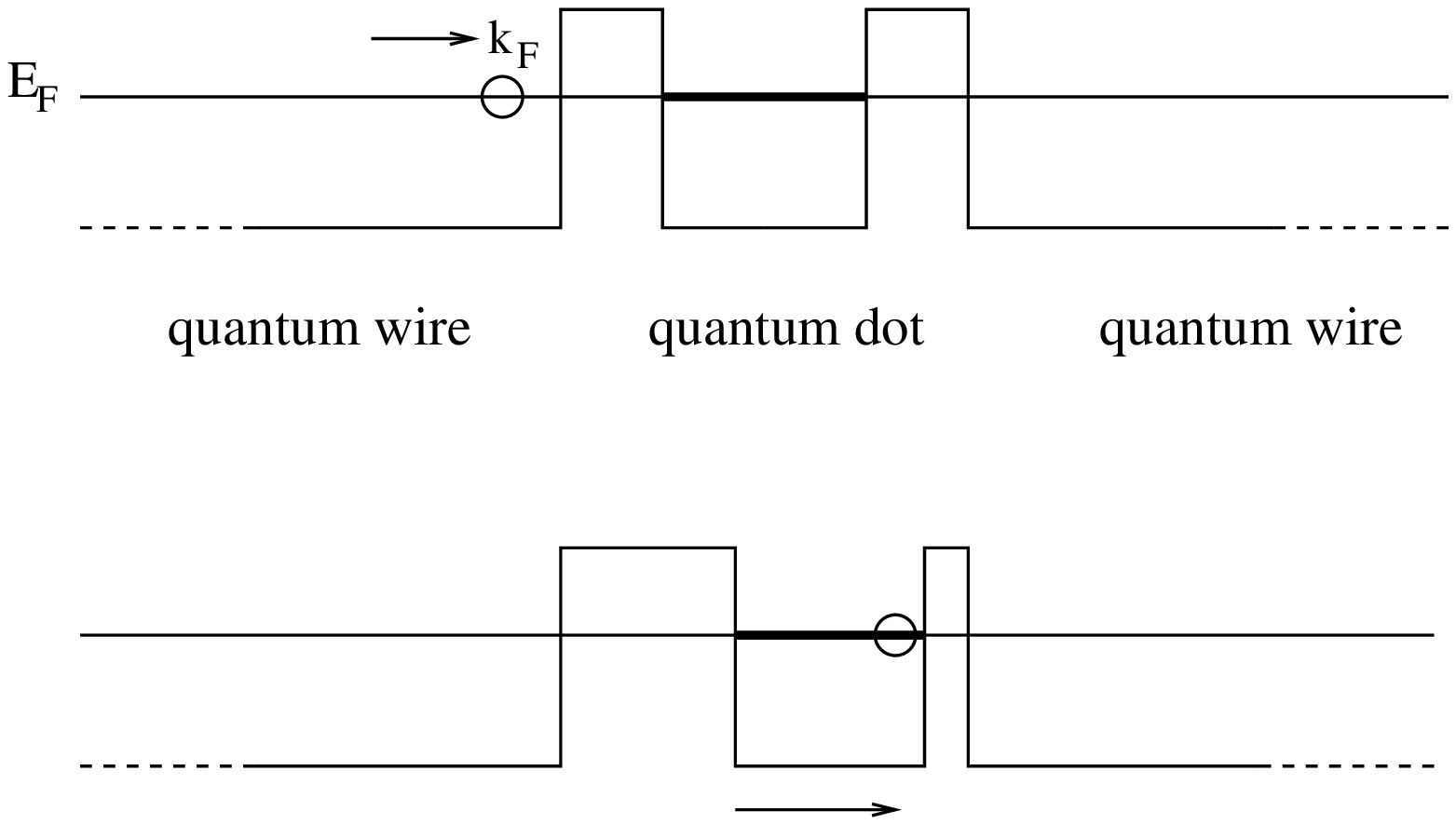}
\caption{}
\end{figure}
\end{center}

\end{document}